# Graphene oxyhydride catalysts in view of spin radical chemistry

**Elena F. Sheka**

*Institute of Physical Researches and Technologies, Peoples' Friendship University of Russia (RUDN University), Miklukho-Maklaya 6, 117198 Moscow, Russia*

**Abstract:** The article discusses carbocatalysis provided with amorphous carbons. The discussion is conducted from the standpoint of the spin chemistry of graphene molecules, in the framework of which the amorphous carbocatalysts are a conglomerate of graphene-oxynitrothiohydride stable radicals presenting the basic structural units (BSUs) of the species. The chemical activity of the BSUs atoms is reliably determined computationally, which allows mapping the distribution of active sites in these molecular catalysts. The presented maps reliably evidence the BSUs radicalization provided with carbon atoms only, the non-terminated edge part of which presents a set of active cites. Spin mapping of carbocatalysts active cites is suggested as the first step towards the spin carbocatalysis of the species.

## 1. Introduction: Modern carbocatalysis and its problems

According to one of the most reputable experts in the field, "catalytic applications of carbon materials are as old as the discipline of physical chemistry, and probably even older" [1]. In fact, amorphous carbon (AC) materials such as activated carbons and carbon blacks have been used for ages in heterogeneous catalysis as either catalysts or catalyst supports. The first documentation of the issue was done about hundred years ago in a report about the aerobic oxidation of oxalic acid occurred on the surface of charcoal [2]. The latest achievements of this biocarbon catalyst can be found in review [3]. By the end of the nineties, researches in the area of AC catalysis were a well-defined field of material science, whose further development is reflected in a large number of publications, thoroughly reviewed in a set of reviews and monographs [4-11]. AC catalysts were discussed with a special emphasis given to the description of their surface features. Simultaneously, the ingrained representation about these carbons as dirty amorphous filtration materials has been overcome and this was replaced with understanding that these materials are built of single or stacked graphene layers [10] thus firmly attributing them to $sp^2$ configured carbon. The graphene layers turned out to be restricted in size to nanometer scale thus putting this $sp^2$ carbon science at the molecular level with graphene-like molecules playing the role of *basic structure units* (BSUs). Evidently, the chemistry of $sp^2$ ACs as well as their catalytic ability is tightly connected with the chemistry of the relevant BSU molecules. The aforementioned ACs are generally handmade and/or industrially produced in massive tonnage. As for natural $sp^2$ ACs, information concerning their reactivity and catalytic ability is much scarcer, however, despite this, quite convincing. This concerns mainly, besides coals, shungite carbons [11-15], for which a decisive role of BSUs for physicochemical processes was clearly shown. Similar investigations of anthraxolites is becoming a top point for the current studies.

Transferring the consideration of ACs catalysis to the atomic-molecular level coincided in time with the emergence of a new branch of catalysis research heralded as metal-free carbon-based catalysis. This was stimulated by the emergence of a sufficient mass amount of nanocarbon materials of new type based on carbon nanotubes (CNTs), graphene oxide (GO), reduced graphene oxide (rGO), and graphene fragments of different origin. The terms "carbocatalysis" and "carbocatalyst" were proposed to describe these carbon-only catalysts [16-20]. It should be commented that the first documentation related to catalytic activity of rGO appeared in 1962, long before the era of $sp^2$ nanocarbons [21]. Confined names AC-Ccats, CNT-Ccats, GO-Ccats, and Gr-Ccats will be used bellow to facilitate the description of AC-, CNT-, GO-, and Gr-based carbocatalysts, respectively. In the latter case, the term combines mainly rGO, better known as "technical graphene" [22] as well as other graphene fragments.

In contrast to ACs, carbocatalysis of CNT- and Gr-Ccats develops extremely rapidly. Annual number of publications obtained from a Scopus search using "graphene and catalysis" as keywords, starting as 74 in 2005, approached 25000 by the end of 2017 [23] and the tendency still remains. Under these conditions, a short list of reviews [24-36] cited in the paper cannot pretend to reflect the state of art in this field in its entirety. However, it has been selectively configured by the author as being directly related to the issues raised in this article. The stormy development of CNT- and Gr-Ccats was stimulated by great hopes for establishing a carbocatalysis mechanism and, as a result, finding ways to improve the already known and develop new catalysts basing on the huge pool of available structural and chemical data. However, the practical situation turned out much complicated. A uniquely large number of studies mentioned above, performed mainly by trial-and-error approach, made it possible to single out only the main tasks of carbocatalysis, the solution of which will help to build an overall skeleton of this scientific and so important practical direction. Aimed at achieving the final goal, practical researches allow now to highlight four nodal braces on which the self-consistent carbocatalysis framework rests. The first concerns detailed knowledge of the structure and chemical content of catalysts themselves. The second deals with understanding which peculiarities of the catalysts structure and chemical content lay the foundation of the catalytic ability. The third is related to the kind of the catalyst-stimulated intermediates must be formed in due course of the wished catalytic procedure. Finally, implementation of the wished reactions accomplishes the long way.

However, such an ideal knowledge-driven four-stage embodiment of the catalytic process is impossible today neither in laboratory, nor in industry. Evidently, there is a considerable progress in each of the fields while many questions remain open. Thus, concerning carbocatalysts structure, the obvious progress is connected with understanding of tight connection of their ability with nanostructuring [11, 23-35], convincingly proven experimentally. Enough to mention the observed enhancement of electrocatalytic activation of oxygen when the size of rGO carbocatalyst decreases [37]. Deeply examined nanostructuring of natural AC-Ccats [38-42] and its great similarity to that of Gr-Ccats found in the course of extended comparative analysis [43-48], not only firmly supports perspective of the AC-Ccats successful catalytic application in future, but also justifies the possibility of a common molecular approach to the consideration of the catalytic ability of both carbocatalyst families. This approach is based on the governing role of the relevant BSU molecules. Actually, real complicated fractal structure of carbocatalysts of both groups can noticeably affect the catalytic activity and this problem is among first rows in the current agenda of interests. Nevertheless, BSU molecules remain the main issues.

The nanostructuring has one more effect on the current carbocatalysis investigation transferring the latter in the field of modern nanotechnology. Once nanotechnological objects, the detailed determination of structural parameters and chemical content of carbocatalysts must be determined using high-tech instruments. However, since each of such instrumentation sees the object from its own viewpoint, the images obtained with different techniques often do not coincide due to which the application of the instrument sets from a broad portfolio of techniques is needed to make a reliable conclusion about the data obtained. The obligatory meeting this requirement is best seen by examining sets of similar samples. So far, such examinations have been performed only twice with respect to CNT-Ccats [49] and a mixture of specially selected synthetic and natural ACs [47]. As shown in both studies, a particular protocol for the characterization of carbon nanostructures must be established to provide a comprehensive characterization of selected samples using common techniques.

When such protocol characterization of the samples is fulfilled and the structural parameters and chemical composition of the relevant BSU molecules are determined, the second stage concerning establishing active sites in the space of the molecules begins. The sites are responsible for adsorption of the reactants, bond breaking and bond formation, as well as desorption of the products. Each of these steps should be confirmed convincingly. However, such an ideal approach has been so far very rare. Usually, the most common is the method of attributing active cites to certain functional groups by analogy. Thus, by the time when the catalytic properties of carbon were discovered, there was a strong view of the relationship between catalysis and either the acidic and basic properties of the catalyst or the decisive role of transition metals in the observed process. The release of carbocatalysts from the metals led to the only mechanism of catalysis associated with the presence of oxygen, sulfur, nitrogen, etc. heteroatoms. This acid-base concept of the active cited, proposed for AC-Ccats more than 30 years ago (see details in review [50]), until recently, has dominated with respect to ACs [1, 5-11] and has retained a strong position relatively to Gr-Ccats as well (see a comprehensive review [51]). In the latter case, the concept has been somewhat corrected emphasizing a particular role of organic molecules due to many of them possessing acidity or basicity as well as redox activity can catalyze reactions. Generally, it has been found that condensed polycyclic aromatic molecules having some functional groups can promote different reactions [52]. However, this analogy between organo- and carbocatalysis cannot explain how to mimic dangling bonds, carbon vacancies, edges and holes of the relevant BSUs with organic molecules. At the same time, carbon vacancies and the periphery of graphene sheets having dangling bonds were believed to be general active sites in many carbocatalytic reactions, such as aerobic oxidations and hydrogenations [20]. Therefore, we are facing a peculiar situation when empirical studies of carbocatalysis evidently respond on the presence of heteroatoms in the relevant catalysts bodies, on the one hand, while there are convincing evidences of a particular role of dangling bonds and defects, on the other. Authors of review [51] suggest a possible solution of the problem in looking for the explanation of this unique property of carbocatalysts by taking into account a joint, delocalized electron pool, which can be modified even by distant atoms. However, until now this approach has not been realized. The unclear situation with active sites does not allow going directly to catalysis itself and raising the question concerning intermediate products of desired catalytic reactions. Until the structure of active sites in BSU molecules is fully clarified and the chemistry of the molecules is understood, the scientific formulation of carbocatalysis is practically impossible and its implementation is destined to obey the trial-and-error practice.

Concluding this short excurse to the modern carbocatalysis science, is necessary to dwell on the role of computational chemistry. If, overall, the virtual fraction of modern graphenics is extremely high (see [53, 54] and references therein), then it is precisely in carbocatalysis that it is unusually modest [24-26, 30]. This is quite understandable due to, first, the general problem concerning active sites, which prevents the creation of reliable models. The greatest that quantum chemists can currently afford is the consideration of BSU molecular models by analogy with well-known examples of organocatalysis [55-57]. However, the chemistry of graphene molecules, spin by nature, significantly differs from that of simpler organic molecules. This peculiarity of graphene molecules is due to a significant correlation between odd electrons of carbon atoms forming honeycomb composition of the molecule. The feature causes a complete delocalization of electron and spin densities [53, 58] that should be taken into account.

The second general problem is connected with the fact that any catalytic reaction is never limited by a single pair of interacting reagents, but includes a set of reacting substances, thus facing investigators with multicomponent kinetics, the balanced result of which is the desired final product. The author knows only two attempts to approach the problem computationally [59, 60]. In the first case, the matter is about large collaborative effort toward an atomic-scale understanding of modern industrial ammonia production over ruthenium catalysts. The success of predicting the outcome of a catalytic reaction from first-principles calculations was provided by a complete clearness of the active sites and rather simple composition of reagents. However, even in this relatively simple case, it took the participation of two dozen people to perform the necessary calculations. In the case of a similar task, involving BSU molecules, the requirements for personal and time resources will increase significantly. The second case concerned carbocatalysis and dealt with oxidation of $H_2S$ on active carbon. Basing on suggested similarity with organocatalysis, active cites of the carbocatalyst were mimicked by quinone and a rectangular (2,5) graphene molecule (as BSU) involving 2 and 5 benzenoid units along armchair and zigzag edges, respectively, dangling bonds of which were terminated. For the first time, spin character of the catalysis was considered, although it was not relatively to the BSU model, but to different spin multiplicities of molecular oxygen and other molecules involved in the process. The study was limited by the computation of energetic parameters of the catalysis elementary steps, which showed a complicated multicomponent character of the catalytic process. As well known, the occurrence of any catalytic reaction is governed by a set of kinetic parameters, among which there are multiple reaction pathways, isomerization, stereo- and regiospecificity related to reactants and products that form free energy basins separated by barriers of different high. Until now, a possibility to quantitatively consider the relevant problems taking into account a large variety of reactants seemed to be impossible. However, the appearance of a new method, called by its authors 'a multi class harmonic linear discriminant analysis' (MC-HLDA) [61], presenting metadynamics with discriminants as a tool for understanding chemistry, inspires great optimism that in future similar complex problems can be solved.

The current paper is devoted to the consideration of active cites in BSU molecules of AC-Ccats, presented by a set of the elitist $sp^2$ ACs of the highest carbonization rank covering both engineered (carbon black) and natural (shungite carbon, anthraxolite, and anthracite) products. Extended detailed analysis of the species structure and chemical content performed earlier [47] allows significantly narrowing the choice of the BSU model structures thus approaching them the reality to the biggest extent. The BSUs are considered in the framework of spin radical chemistry of graphene [53, 58], in view of which the species represent molecular stable radicals, spin density of which is delocalized over all carbon atoms of the molecules with the highest values at edge atoms. The role of heteroatoms in the BSU structures is discussed in the light of their influence on the radical properties of the molecules. Calculations were performed in the framework of UHF semi empirical AM1 approach (technical details can be found in [47]). The data, part of which were earlier obtained and discussed [47], are complemented by new ones related not only to graphene oxyhydrides, but also to the latter, additionally decorated with nitrogen and sulfur atoms, under the common name graphene oxynitrothiohydrides.

## 2. Structure and chemical composition of amorphous carbons BSUs

From the beginning of the graphene era, there has been an opinion that chemical derivatives of graphene, mainly GO, rGO and teflon, came into life together with graphene itself. The first breakdown was caused by the establishment that natural AC, known as shungite carbon, is a natural deposit containing millions tons of rGO [38]. Then similar conclusions were made about the natural anthraxolite [42] and anthracite [48]. In time, it came to the industrial multi-tonnage carbon blacks as well [47]. A recent extended study was devoted to a detailed investigation of the structure and chemical composition of a set of these ACs of the highest carbonization [47]. The performed analytical study has shown that all the studied ACs are multielemental compounds with dominating contribution of carbon while complemented with different minorities. Among the latter, oxygen plays the main role constituting a few wt% in all samples. Hydrogen is the next and is of high importance for carbon materials. However, the hydrogen weight content is comparable with that of other impurities involving sulfur, nitrogen, chlorine, silicon, and different metals. Therefore, known more than one-and-half thousands of years, AC is undergoing a rebirth and appears as a set of agglomerative compositions of framed graphene oxynitrothiohydride (GONSH) molecules, thus becoming a special subject of modern nanotechnology.

Common to the studied amorphics are the principles of the formation of their architecture. Based on multilevel fractal composition of shungite carbon [39], the latter rests on the BSUs as structure elements of the first level that are composed in stacks at the second level with grouping the stacks in globules at the third level and completing the agglomeration of globules at the fourth level. The first two levels are directly supported with detailed HRTEM studies as well as neutron and X-ray diffraction [47]. The last two levels correlates perfectly with multidimensional porous structure of the species [62-64]. In the current paper these and other commonalities of $sp^2$ ACs will be exemplified for three natural and one synthetic ACs presented by shungite carbon (ShC), anthraxolite (AnthX), and anthracite (AnthC) in the first case and by carbon black 699632 (CB632) produced by Merk-Sigma-Aldrich [65]. Necessary information concerning their structural and chemical content is listed in Tables 1 and 2. All the species are nanostructured and mesoporous. A general view on their structures is presented in Fig. 1. As seen in the figure, natural amorphics and carbon black represent complexes of micrometer aggregates with a lot of free space between them exhibiting the origin of the species porous structure. A close similarity of the structure appearance of all samples confirms a common structural architecture for the species. Structural parameters related to BSU themselves ($L_a$) and stacks of them ($L_c$) are given in Table 1. Both values are of nanometer scale thus providing nanostructuring of the samples. In the recent study [47], it was possible to propose for the first time model BSUs structures corresponding to the data given in the tables, concerning only C-O-H triads. The basis of all these graphene oxyhydrides (GOH) models was the (5, 5)NGr molecule, representing a rectangle graphene sheet with five benzenoid units along both armchair and zigzag edges, commensurate with the real BSUs in the parameter $L_a$. The choice allows drawing the main attention to the framing areas of the models without distracting it to the structure of the carbon core. Implementing the determined chemical content [47], proposed GOH BSUs are described by general formula each, that are $C_{66}O_4H_6$ (ShC),

C$_{66}$O$_4$H$_{10}$ (AnthX), C$_{66}$O$_5$H$_{12}$ (AnthC), and C$_{66}$O$_4$H$_3$ (CB632). Obviously, each of these formulae covers a large set of GOH molecules differing by chemical bonding of oxygen and hydrogen atoms with the parent molecule. The bonding types were disclosed by scrupulous XPS study based on the modern multi-complex-oxygen-containing-group concept [66] that revealed their different configurations for the species [47]. According to multitarget character of the spin chemistry of graphene molecules [53, 58], implementation of each chemical formula given above can be achieved by different ways due to which a large number of GOH molecules can be suggested for each amorphics. Thus, first six- (CB632) and twelve- (ShC and AnthX) member sets were proposed for each of them [47]. Schematic presentation of a shungite carbon globe, presented in Fig. 1, is constructed of a mixture of 4-7 layer stacks composed of different members of the set of models discussed there. Not-carbon atoms are especially highlighted to draw attention how colorful are the purest carbon materials.

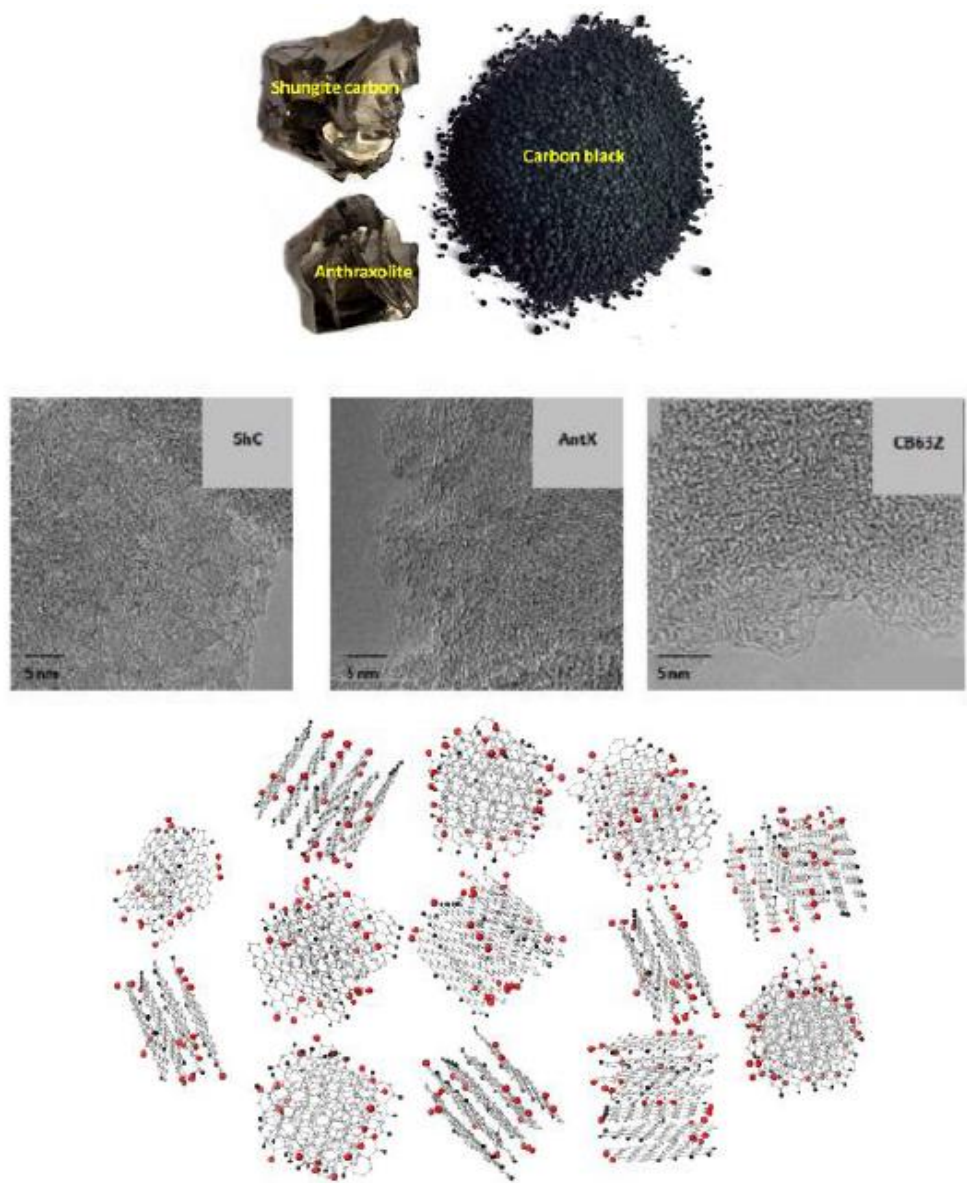

**Figure 1.** Appearance (top) and TEM image (middle) of natural amorphous carbon. Bottom: Planar view on a globule of shungite carbon consisting of four-, five-, six, and seven-layer stacks. Linear dimension of the globe is of ~6 nm. The chemical composition of all the BSUs is described by the formula C$_{66}$O$_4$H$_6$. Light gray, black and red balls depict carbon, hydrogen, and oxygen atoms, respectively.

Table 1. Structural parameters of amorphous carbons [47] [1]

| Samples | d (Å) | $L^c_{CSR}$, nm | Number of BSU layers | $L^a_{CSR}$, nm |
|---|---|---|---|---|
| Graphite | 3.35 | >20[2] | ~100 | >20 |
| ShC | 3.47(n); 3.48(X) | 2,5(n); 2.0(X) | 7(n); 5-6(X) | 2.1(X) |
| AnthX | 3.47(n); 3.47(X) | 2.5(n); 1.9(X) | 7(n); 5-6(X) | 1.6(X) |
| AnthC[3] | 3.50(X) | 2.2(X) | 5-6(X) | 2.1(X) |
| CB632 | 3.57(n); 3.58(X) | 2.2(n); 1.6(X) | 6(n); 4-5(X) | 1.4(X) |

[1] Notations (n) and (X) indicate data obtained by neutron and X-ray diffraction, respectively.
[2] The definition ">20 nm" marks the low limit of the dimension pointing that it is bigger than the coherent scattering length of crystalline graphite equal to ~20 nm along both *a* and c directions. Actual dimensions are of micrometer range.
[3] The data are taken from [48].

Table 2. Chemical content of amorphous carbons [47]

| Samples | Elemental analysis, wt% | | | | | XPS analysis, at% | | |
|---|---|---|---|---|---|---|---|---|
| | C | H | N | O | S | C | O | Minor impurities |
| ShC | 94.44 | 0.63 | 0.88 | 4.28 | 1.11 | 92.05 | 6.73 | **S** - 0.92; **Si** – 0.20; **N**-0.10 |
| AnthX | 94.01 | 1.11 | 0.86 | 2.66 | 1.36 | 92.83 | 6.00 | **S** - 0.85; **Si** – 0.25; **N**-0.07 |
| AnthC[1] | 90.53 | 1.43 | 0.74 | 6.44 | 0.89 | 92.94 | 6.61 | **Cl** - 0.11 - **S**: 0.34 |
| CB632 | 97.94 | 0.32 | 0.04 | 1.66 | 0.68 | 93.32 | 6.02 | **Si** – 0.66 |

[1] The data are taken from [48].

## 3. Radical character of the amorphics BSUs; active cites of the carbocatalysts

*3.1. Graphene oxyhydrides*

For the first time, the question of the relationship between the ACs catalytic ability and the particular properties of its BSUs was raised in [47]. As was shown for GOH models, BSUs are radicals with quite high chemical reactivity. Figure 2 presents a comparative view on the radical behavior of the studied amorphics exemplified by 'chemical portraits' of the relevant model GOHs. The portraits are presented by the image pairs, left of which are the distributions of the effectively unpaired electrons $N_{DA}$ ($N_{DA}$ maps of atomic chemical susceptibility (ACS) [52, 53]) over the GOH molecules atoms while the right part are images of the equilibrium atomic structures of the species. Each of the exhibited molecules heads sets of the GOH models suggested for the studied amorphics [47]. The difference in the chemical bonding of the models was discussed there in details as well.

The $N_{DA}$ map of the parent (5, 5)NGr molecule is shown on the figure top. Bright spots on all the maps indicate cites of the most chemical reactivity. Quantitative presentation of the reactivity is shown by ACS plottings over carbon atoms of the GOH molecules in Fig. 3. The plottings are given on the background of that one belonging to the (5, 5)NGr molecule which is presented in the Z→A format on the top of Fig. 2. As seen in the figure, large $N_{DA}$ values of 0.9÷1.3 e distinguish 22 edge atoms while values ≤ 0.3 e belong to basal-plane atoms. This characteristic feature of naked graphene molecule, repeatedly confirmed experimentally (see [53] and references therein), as a whole is preserved in the GOH molecules as well but is evidently disturbed. As visible in Figs. 2 and 3, the inclusion of hydrogen and oxygen into the parent molecule circumference expectedly inhibits the activity of the edge carbon atoms, directly involved in the new bonding. At the same time, the figures show enhancement of the activity of basal-plane atoms. The feature clearly demonstrates once again a peculiar collective character of chemical events occurring with graphene molecules [58]. The molecular chemical susceptibility $N_D$ [53] of the parent molecule (33.49 e) considerably reduces in all the GOH molecules and constitutes 23.65 e ($C_{66}O_4H_6$), 22.23 e ($C_{66}O_4H_{10}$), 20.49 e ($C_{66}O_5H_{12}$), and 29.66 e ($C_{66}O_4H_3$). The $N_D$ values of graphene molecules as well as their distributions over atoms in terms of $N_{DA}$ are governed by the C=C bond length distribution [67] that is evidently significantly disturbed in the basal plane by addition of the discussed minorities in the GOH circumference.

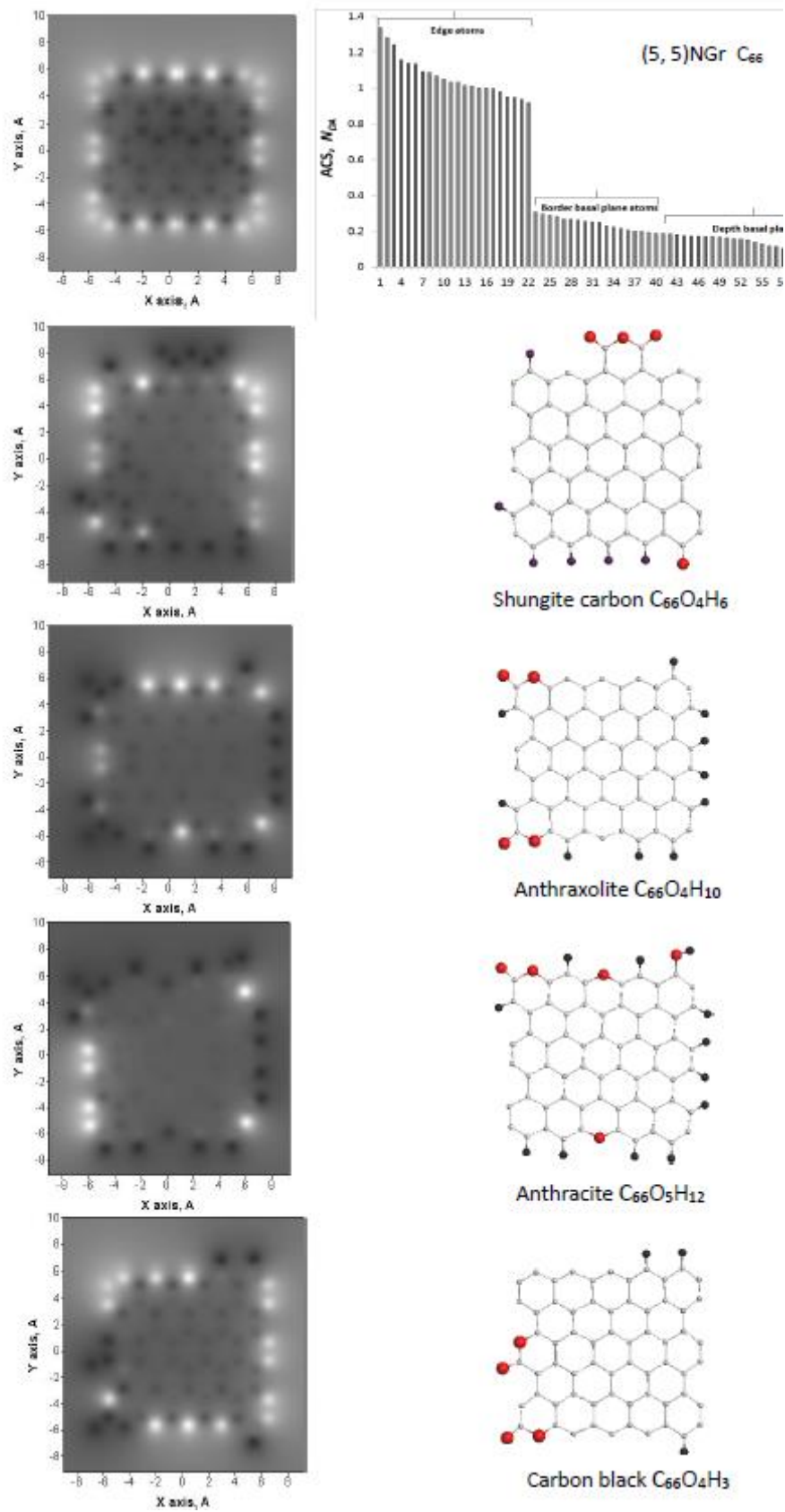

**Figure 2.** ACS $N_{DA}$ maps (left) and equilibrium atomic structures (right) of GOH models of amorphous carbons. Top: ACS map and Z→A $N_{DA}$ distribution over atoms of the parent (5, 5)NGr molecule.

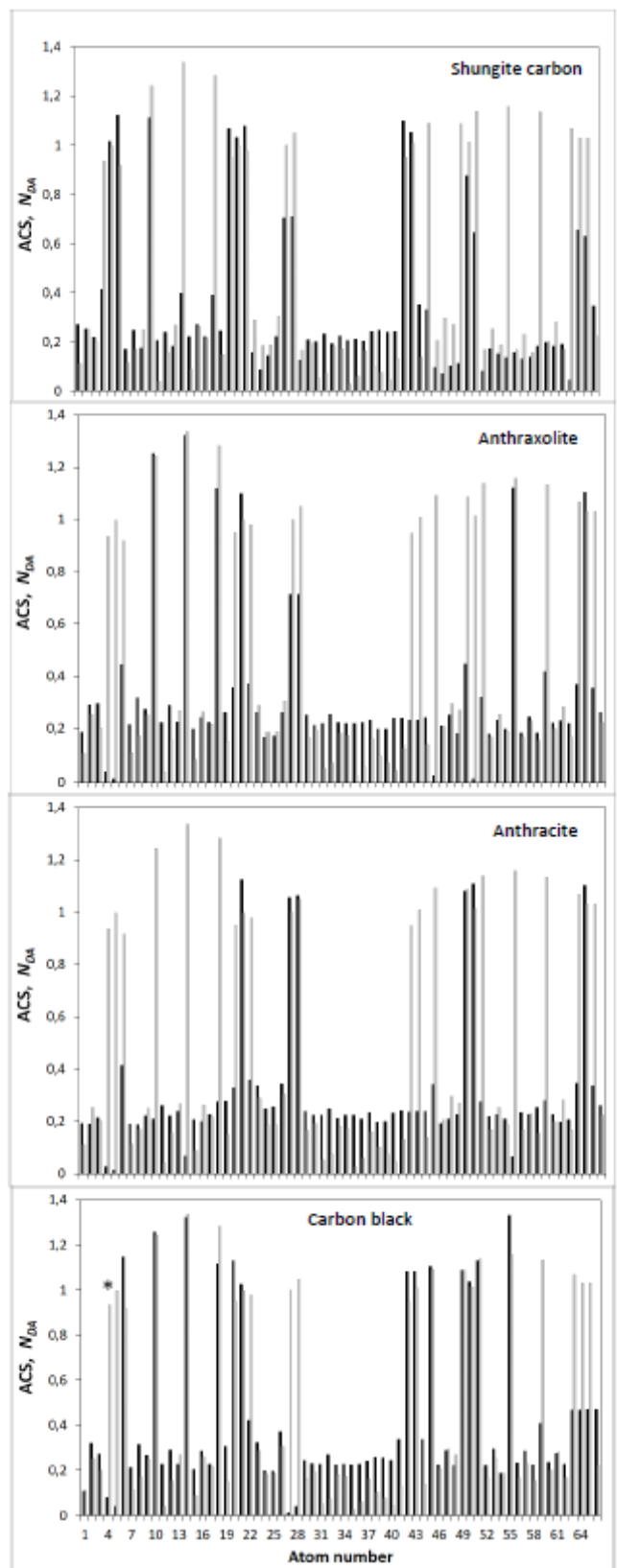

**Figure 3**. ACS $N_{DA}$ distribution over carbon atoms of the (5, 5)NGr (light-gray) and GOH (black) molecules.

### 3.2. Graphene oxynitrothiohydrides

As follows from Table 2, the discussed GOH models do not reproduce the real BSUs completely, which evidently include sulfur and nitrogen. Data presented in the table show that the amount of nitrogen in the above GOH molecules constitutes ~0.5 atom, which means that, on the average, only half of the (5, 5)NGr-based molecules involve one nitrogen atom in their structure. As for sulfur, the element content corresponds to one atom per three molecules of GOH molecules of shungite carbon and anthraxolite, per four molecules of anthracite and per five molecules of carbon black. It should be remembered that CHNS and XPS analysis are related to different parts of the BSU molecules. Thus, the former 'sees' the molecule as a whole while the latter deals with the molecule edge only [47]. Therefore, according to the data in Table 2, nitrogen atoms are located in the GOH molecules basal plane while sulfur atoms in the relation of ~1:1 can be located either in the molecules plane or in the circumference. Considering the features, let us analyze maximum changes in the chemical portraits of GOH caused by the molecules decoration with nitrogen and sulfur atoms. Two GOH molecules $C_{66}O_4H_6$ and $C_{66}O_4H_3$ related to shungite carbon and carbon black, respectively, were chosen for this comparative study. When constructing the molecules, we proceeded from the following facts. As said earlier, a single nitrogen atom is located in the above molecules basal plane. Since, according to Table 1, the interlayer distance between neighboring BSUs molecules is close to that in graphite, the only way for nitrogen to exist in the basal plane is its intrusion into the benzoid cycle forming a pyridine-like structure. The same concerns the fraction of sulfur atoms that should be embedded in basal plane. Suggested N and S configurations of the studied GOHs are widely discussed with respect to graphene-based carbocatalists [68-70]. The carbon targets to be substituted with either N or S atoms were selected among border basal plane atoms (see top ACS histogram in Fig.2). The corresponding structures of GONSH ShC_N and CB_N as well as ShC_NS1 and CB_NS1 are shown in Figs. 4 and 5. Since the formation of thiophene-like structure on the edge of graphene sheets is highly probable [68], this construction led the foundation of GONSH ShC_NS2 and CB_NS2 models shown in the figures. Image ACS maps in the figures present detailed vision of the influence of additional decoration on the GOH molecules chemical activity with single atoms of nitrogen and sulfur. Qualitatively, the effect can be traced by the relevant plottings shown in Fig. 6. As is visible in the figures, each decoration act causes remarkable changes in the ACS distribution thus touching the C=C bond distribution of the whole molecules. However, despite the effect in each case is expectedly individual, the general pattern of the ACS distribution is conserved. Before to proceed further with linking the ACS peculiarity with catalytic ability of amorphous carbons we must verify the correctness and reliability of the proposed models. For this, consider the structural features of the proposed molecules and stacks of them.

Experimentally proven is the fact that BSUs molecules form stacks consisting of 4-7 units with the average interlayer distance $d$ close to that of graphite (see Table 1). The fact strongly evidences practically flat structure of the molecules. As for model structures discussed in the current paper, GOH molecules shown in Fig. 2 are flat. Once tightly packed and decorated with hydrogens and oxygens, vdW radii (vdWR) of which are less than that of carbon, they are separated from each other at a distance determined by the doubled carbon atom vdWR of 1.71 Å [71]. In contrast, intrusion of sulfur atom in the circumference area causes enlargement of the interlayer distance even when the molecules are flat due to significantly bigger atom vdWR (1.84 Å [71]). Therefore, the presence of sulfur atoms in the BSUs circumference can explain experimentally observed distance excess on 0.25 Å over that for graphite. However, this is not the only reason of the enlargement observed experimentally. The other is connected with the reconstruction of the basal plane of the molecules after the intrusion of either nitrogen or sulfur atom. As was found in the current study, the plane disturbance depends on the position of the target carbon atom substituted with heteroatom. The effect is expectedly more significant for sulfur and is as bigger as larger the distance of the target atom from the molecule edges. Figure 7 shows the maximum effect to be occurred for the studied molecules. The flatness of GONSH molecules ShC_NS3 and CB_NS3 is drastically disturbed, which can lead to an increase in the interlayer distance *d* by more than 1.5 A. However, in practice this did not happen. Thus, in order to reduce such a large discrepancy between the observed and characteristic *d* values of the last two molecules, we have to accept that both basal-plane nitrogen and sulfur atoms are located in the area formed by border-basal-plane carbon atoms. Molecules ShC_S1 and CB_S1 present just the case and together with the other two pairs of molecules shown in Figs. 4 and 5 form the ground for the discussion of catalytic ability of shungite carbon and carbon black.

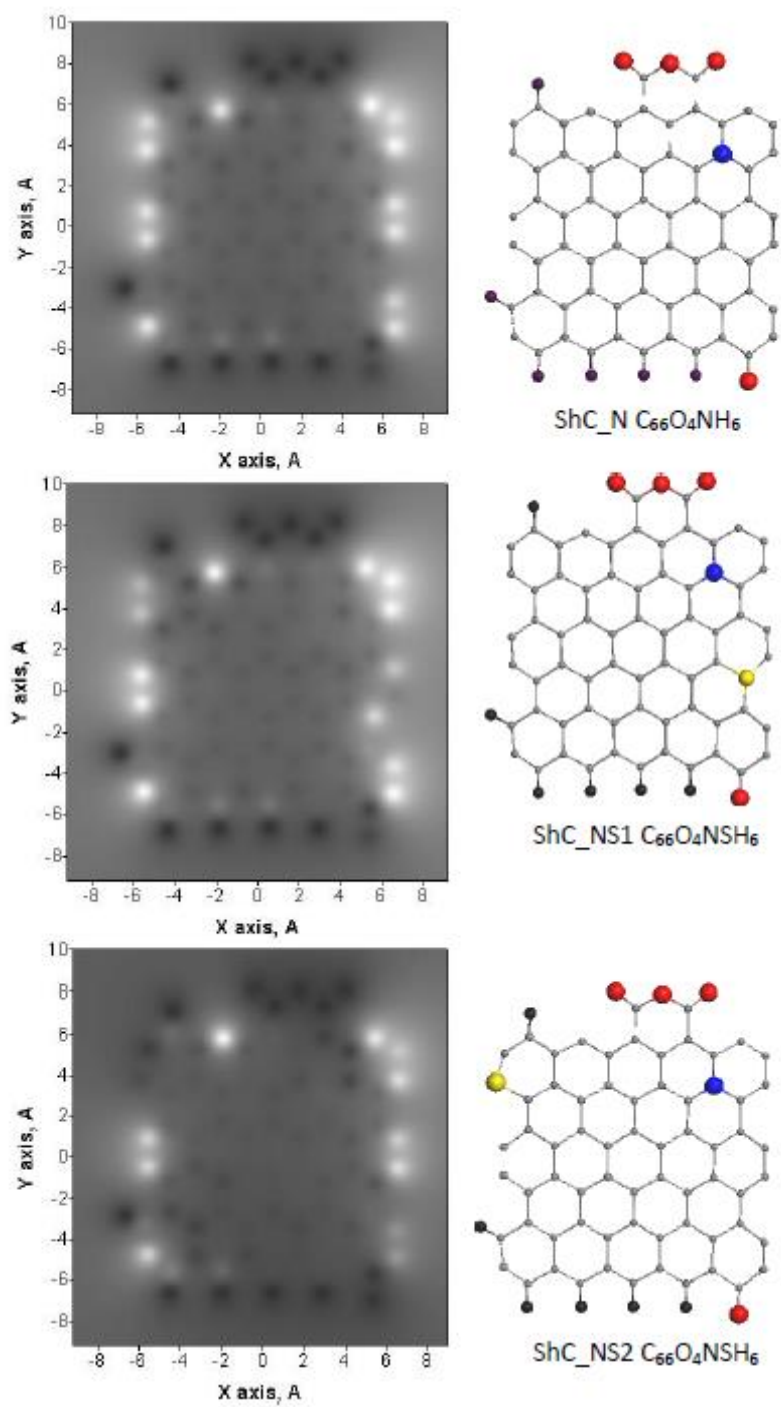

**Figure 4.** ACS $N_{DA}$ maps (left) and equilibrium atomic structures (right) of shungite carbon GONSH model decorated by single atoms of nitrogen (dark blue balls) and sulfur (yellow balls).

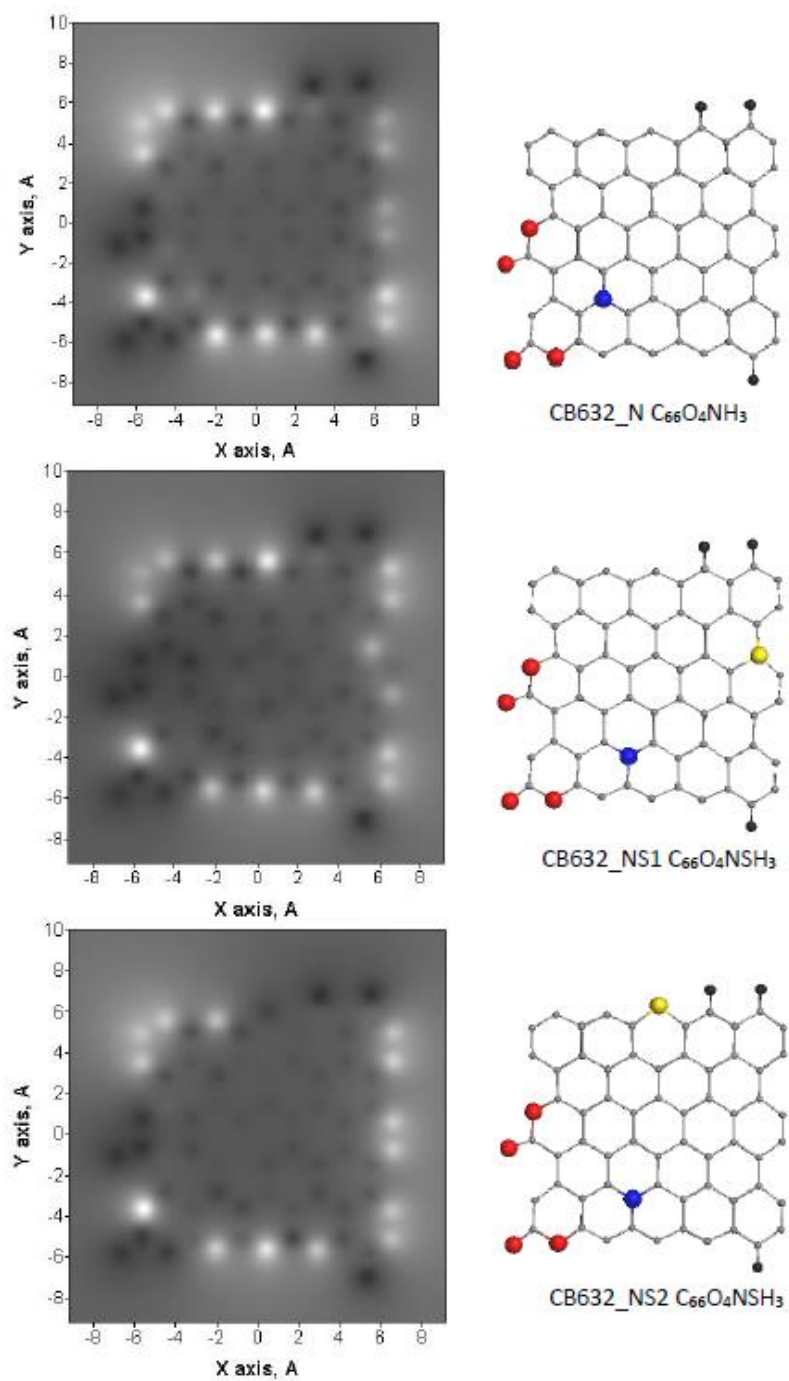

**Figure 5.** ACS $N_{DA}$ maps (left) and equilibrium atomic structures (right) of carbon black GONSH model decorated by single atoms of nitrogen (dark blue balls) and sulfur (yellow balls).

*3.3. Active sides of amorphous carbocatalysts*

    Thus, the mixture of GOH and GONSH molecules (of different types such as GOH_N, GOH_S, and GOH_NS, in the latter case) forms the pool of BSUs molecules of the studied AC-Ccats. Evidently, the data on the ACS $N_{DA}$ distributions over these molecules atoms lay the foundation of answering the main question of the molecules catalytic ability concerning the active cites of the studied carbocatalysts. Looking at the ACS image maps in Figs. 2, 4, and 5 as well as at the ACS histograms in Figs. 3 and 6, one can get quite strait and firm answer: remaining non-determined edge carbon atoms of BSU molecules are the cites under question. These atoms fraction constitutes from ~1 to 2 % in the studied carbocatalysts, which

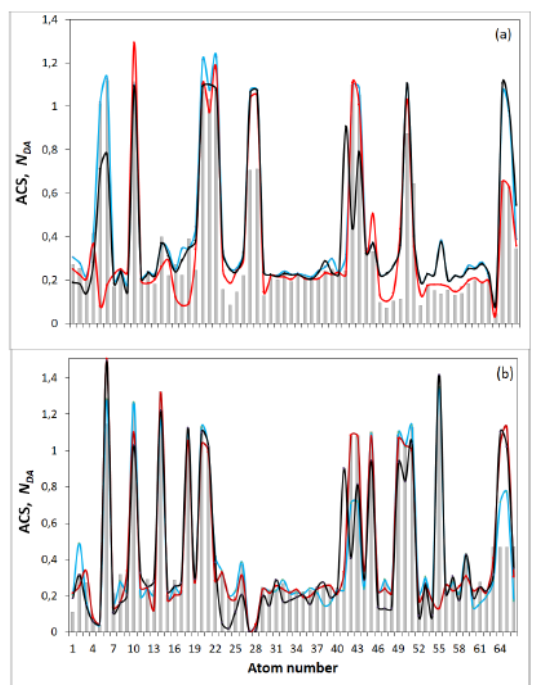

**Figure 6.** ACS $N_{DA}$ distribution over carbon atoms of the $C_{66}O_4H_6$ (a) and $C_{66}O_4H_6$ (b) before (histograms) and after (curves) decoration by single N atoms (light blue curves) and additionally by single S atoms in the circumference (red curves) and basal plane (black curves).

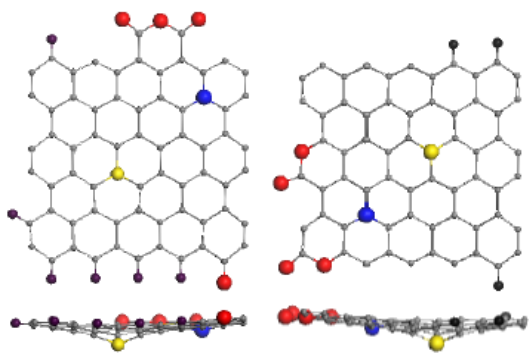

**Figure 7.** Top and side views of equilibrium structure of ShC_NS3 ($C_{66}O_4NSH_6$) (left) and CB632_NS3 ($C_{66}O_4NSH_3$) (right) GONSH molecules (see text).

explains low yield of carbocatalytic reactions. Additionally, a comparative analysis of all the data obtained shows that chemical reactivity of the studied ACs carbocatalysts gradually reduces when going from carbon black to shungite carbon, anthraxolite, and anthracite. The sequence well correlates with experimental observations.

As for functional groups involving heteroatoms, a dominant role of which in carbocatalysis has been largely discussed until now, as seen in Figs.4-6, they completely lose their activity after attaching the carbon core and are not capable to further lead any chemical reaction while neighbor 'empty' carbon atoms are ready to play the role. The feature is a straight consequence of the radical essence of the relevant molecular catalysts and is characteristic for $sp^2$- configured fragments of honeycomb structure of framed graphene derivatives. In the practice of organocatalysis, molecules of this type are quite rare. This circumstance becomes a serious obstacle to the validity of using analogical empirical data of organocatalysis to explain or predict the effect of carbocatalysts. At the same time, the role of heteroatoms in carbocatalysis is very important. According to the author viewpoint, they provide stabilization of molecular carbocatalysts, preserving their radical properties over long time intervals. Apparently, the roots of this stabilization are of kinetic origin. It is possible that the presence of heteroatoms in the circumference of BSUs leads to the appearance of large barriers to the additive reactions involving neighboring non-terminated carbon atoms under ordinary conditions. As a result, the catalytic activity of BSUs is possible only with a significant violation of these conditions and/or with the participation of certain reagents that reduce or remove these barriers. The consequence of this may be a rigid selection of possible catalytic reactions involving carbocatalysts. These and similar issues, pending resolution, appear one after another on the agenda of modern carbocatalysis.

Computational examination of graphene molecules is exclusively large while the fact of their radical nature has not been usually taken into account. Ignoring the fact (in contrast to the case of legitimate small stable radicals [72]) should be connected with the way of progress in the description of their electron and spin properties in the framework of computational quantum chemistry (QCh). At this point, the considerations of small stable radicals and graphene molecules proceeded in different directions. The former, once small by size, were considered by using high level QCh approaches, mainly based on Hartree-Fock approximation, that take radical properties into account [67]. The latter, once large, were considered by using much simpler approaches, predominantly various versions of DFT. However, this approach is spin inactive and losses radicals substituting them by ordinary closed-shell molecules (see a profound discussion of the issue in [53, 54]). At the same time, the dictate of DFT-based virtual QCh, particularly strengthened by the current graphene science is very strong, which does not allow raising the curtain and bringing the main actor of catalysis - stable radicals to the scene. Against this background, a study [73] that voiced the term "radical" in relation to $sp^2$ amorphics for the first time when attempting to explain carbonaceous soot inception and growth in term of resonance-stabilized hydrocarbonradical chain reactions, is quite revolutionary. The presented viewpoint on the carbocatalyst models as stable molecular radicals drastically changes a general concept of carbocatalysis and allows making a conscious choice of the catalytic reactions in advance thus using the multi-ton mass of the existing AC carbocatalysts for their implementation.

## 4. Conclusion

Catalysis has always been and remains the most difficult part of synthetic chemistry. The mechanism of catalysis, determined by a large number of thermodynamic and kinetic factors, is largely incomprehensible theoretically to this day as well. Its new manifestation in the form of carbocatalysis, consisting according to the modern representation of three parts: CNT-, Gr-, and AC-Ccats, not only does not add simplicity to the understanding of this complex process, but also, on the contrary, significantly complicates available knowledge by introducing spin into its everyday life. The participation of the electron spin in catalysis is as old as catalysis itself. It is enough to recall all the catalytic oxidation reactions involving triplet molecular oxygen and the presence of spin in most transition metal catalysts. However, the spin presence in these cases was amenable to standardization by introducing empirical parameters. In the case of carbocatalysis, spin is introduced into this process on an ongoing basis. Concomitant from agglomerates of molecular stable radicals, carbocatalysts are characterized by delocalized spin density, labile and easily amenable to external influences, not subject to standardization, thus making the spin consideration permanent and obligatory. Meeting this requirement has serious consequences. In the field of empirical catalysis, the spin factor greatly complicates the perception of carbocatalysis by analogy with organocatalysis. Organocatalysts are, in most cases, molecules that do not have radical properties. On the other hand, computational carbocatalysis requires taking into account the correlation of odd electrons of carbon atoms, and, therefore, the use of high-level computational methods or, at least, the unrestricted Hartree-Fock one, which significantly discounts the use of simplified methods such as DFT, tight binding, restricted Hartree-Fock and others. Therefore, it is obvious that spin carbocatalysis is a thing of the distant future. Nevertheless, the aim of this work was to draw attention to the fact that even today taking spin in service can help to notice those characteristic signs, in both empirical practice of carbocatalysis and first attempts to interpret it theoretically, that will allow to confidently construct the building of spin carbocatalysis in the future.

**Acknowledgements.** The publication has been prepared with the support of the "RUDN University Program 5-100".